\documentclass[conference]{IEEEtran}
\IEEEoverridecommandlockouts
\usepackage{cite}
\usepackage{amsmath,amssymb,amsfonts}
\usepackage{algorithmic}
\usepackage{graphicx}
\usepackage{textcomp}
\usepackage{xcolor}
\def\BibTeX{{\rm B\kern-.05em{\sc i\kern-.025em b}\kern-.08em
    T\kern-.1667em\lower.7ex\hbox{E}\kern-.125emX}}

\makeatletter
\newcommand{\linebreakand}{%
\end{@IEEEauthorhalign}
\hfill\mbox{}\par
\mbox{}\hfill\begin{@IEEEauthorhalign}
}
\makeatother

\begin{document}

\title{Quantum Natural Language Processing based Sentiment Analysis using lambeq Toolkit\\
}

\author{\IEEEauthorblockN{ Srinjoy Ganguly }
\IEEEauthorblockA{\textit{Escuela Técnica Superior de
		Ingeniería de Sistemas Informáticos } \\
\textit{Universidad Politécnica
	de Madrid}\\
Madrid, Spain \\
srinjoyganguly@gmail.com}
\and
\IEEEauthorblockN{ Sai Nandan Morapakula}
\IEEEauthorblockA{\textit{Electrical and Electronics Engineering } \\
\textit{Karunya Institute of Technology and Sciences }\\
Coimbatore, India \\
sainandanm2002@gmail.com}
\linebreakand
\IEEEauthorblockN{Luis Miguel Pozo Coronado}
\IEEEauthorblockA{\textit{Escuela Técnica Superior de
		Ingeniería de Sistemas Informáticos } \\
\textit{Universidad Politécnica
	de Madrid }\\
Madrid, Spain \\
lm.pozo@upm.es}
}

\maketitle

\begin{abstract}
Sentiment classification is one the best use case of classical natural language processing (NLP) where we can witness its power in various daily life domains such as banking, business and marketing industry. We already know how classical AI and machine learning can change and improve technology. Quantum natural language processing (QNLP) is a young and gradually emerging technology which has the potential to provide quantum advantage for NLP tasks. In this paper we show the first application of QNLP for sentiment analysis and achieve perfect test set accuracy for three different kinds of simulations and a decent accuracy for experiments ran on a noisy quantum device. We utilize the lambeq QNLP toolkit and $t|ket>$ by Cambridge Quantum (Quantinuum) to bring out the results. 
\end{abstract}


\section{Introduction}
Taking computational speeds and performance into consideration, quantum computers are exponentially faster than the present generation classical computers. Until the last two decades or the end of the $20^{th}$ century quantum computer was a fictional story developed by great mathematicians and physicists such as Richard Feynman, Erwin Schrodinger, David Deutsch, etc. In the early $21^{st}$ century quantum computers gained its importance and the fictional story was greatly read, understood and was theoretically well developed. Recently, the first quantum computers have been built, some of them have been made publicly available, and they have already gained its significance and showed its power in various domains like machine learning, chemistry, natural language processing, biomedicine, etc. In this paper we will see how quantum computers can help us in improving the domain of natural language processing. 

Quantum computing is itself a nascent field so is Quantum Natural Language Processing (QNLP), we take the phenomenon of superposition, entanglement, interference to our own advantage and run NLP models or language related tasks on the hardware. As of now we are currently in the Noisy Intermediate-Scale Quantum (NISQ) \cite{Preskill_2018} computers era, where the error rate is directly proportional to the number of qubits and information a qubit contains can be lost easily which is why quantum computers are stored at very cool temperatures and requires great maintainence.

QNLP is different from classical NLP. QNLP has its origins in abstract mathematical theory which includes category theory - especially monoidal categories, diagramatic quantum theory and ZX calculus. To gain more understanding about the concepts of diagrammatic quantum theory, the reader can refer to \cite{coecke_kissinger_2017} which explains the fundamentals of diagrammatic reasoning for quantum theory and is the core of QNLP. Since a model of natural language is being equivalent to a model which explains quantum mechanical phenomena, this approach makes QNLP quantum-native. By this process linguistic structure can be encoded easily where as encoding grammer in classical is very costly.

In this paper we are going to see how accurately quantum computers can predict the sentiments, where they are already trained with around 130 sentences. We will also see how classical computers and quantum computers with embedded noise will give the results and compare them to get a better understanding of why we need quantum computers and how powerful they are. 

The rest of the paper is ordered as follows: section 2 gives an introduction to the related work that is done and the research going on in this field; section 3 gives a clear picture and brief intuition on QNLP and also explains the sentiment classification experiment; In section 4 we discuss the results of classical, quantum and quantum with noise devices, section 5 summarises the work and also proposes some future lines of work in the domain of QNLP.

\section{Related Work}

As researchers, scientists, enthusiasts identified the capability of quantum devices and the power they have got, more effort and time is put into this domain. Despite of QNLP being a new and emerging field, Noisy Intermediate Scale Quantum(NISQ) devices have already led to some propitious results \cite{Zeng_2016} and also appiled to divergent field such as quantum music \cite{miranda2021quantum}.

As we all know that the applications of classical NLP is already seen in our day-to-day lives. Voice assistants such as Siri and Alexa are the best examples. The problem and constraint with classical NLP is that it can only read and decipher bits but cannot deeply understand the meaning of the language and that is where there is scope for quantum to do this in a meaning-aware manner. 

In QNLP, the sentences are depicted by variational quantum circuits and each and every word in a sentence is transformed into quantum states by using parameterized quantum gates. With the help of the variational circuit technique, QNLP becomes NISQ-friendly \cite{coecke2020foundations}.  

Scientists and researchers at Cambridge Quantum developed the first high level python framework for QNLP named lambeq. This unique toolkit is a open source package which offers the functionality of converting sentences into quantum circuits \cite{kartsaklis2021lambeq}.

The first and foremost effort of designing and execution of natural language models on a real quantum computer was accomplished by Cambridge Quantum where they used an intermediate dataset containing  sentences of two different classes - Food or IT and results obtained were profound as given in \cite{lorenz2021qnlp}. 

In this work, we present the first application of QNLP for sentiment analysis on an intermediate level dataset where we achieve successful results in a binary classification of sentiments - positive and negative sentences. We demonstrate that for both classical and quantum simulations we achieve proper convergence. 

\section{Methodology and Experiments}

We have utilized the Distributional Compositonal Categorical (DisCoCat) \cite{coecke2010mathematical} framework for our task of sentiment classification. The DisCoCat framework provides a unique way of combining different constitutent words to form the meaning of a whole sentence. It follows a compositional mechanism for the grammar to entangle word meanings which are distributed in a vector space. Lambek's pregroup grammar is used in DisCoCat to retrieve the meaning of a sentence from the meaning of the words. 

\subsection{Compositional Model of Meaning}\label{Discocatmodel}

The compositional model of meaning is inspired from Lambek's pregroup grammar. In this formalism, we assign atomic types 'n' for noun and 's' for sentence to the words present in a sentence which assists in the composition of the word meanings together. The grammatical rules to compose different types of sentences can be found in \cite{lambek2008word} and the string diagrams shown have utilized those rules given by Lambek.

\begin{figure}[htbp]
\centerline{\includegraphics[scale = 0.15]{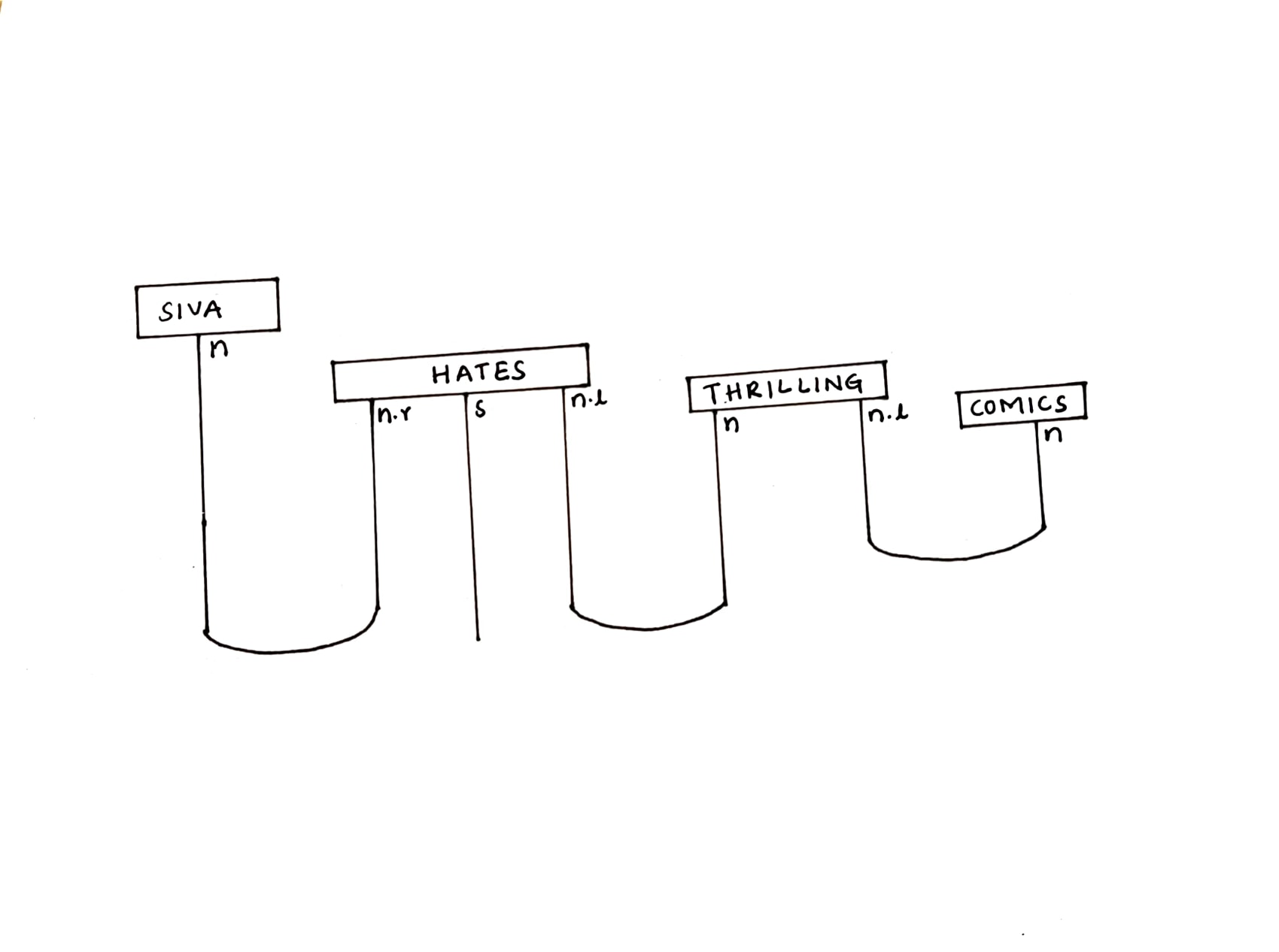}}
\caption{String diagram of "siva hates thrilling comics".}
\label{fig1}
\end{figure}

Fig.~\ref{fig1} shows the string diagram for a sentence "siva hates thrilling comics" where the words are represented by boxes or alternatively triangle shaped boxes and the wires (cups and straight wires) represent the entagling effect which composes the words together to give the meaning of the sentence, which is the grammar. The juxtaposition of the atomic types reduces to 's' which signifies that the sentence is grammatically correct. This juxtaposition is solved in \eqref{eq1}.
\begin{equation} 
n \cdot n^r \cdot s \cdot n^l \cdot n \cdot n^l \cdot n \rightarrow 1 \cdot s \cdot 1 \cdot 1 \rightarrow s  \label{eq1}
\end{equation}

Due to today's NISQ devices i.e. small number of qubits, qubit decoherece and since the string diagrams themselves are resource intensive, they need to be rewritten into a more NISQ friendly version which is going to consume less number of qubits to represent the sentences. 

\begin{figure}[htbp]
\centerline{\includegraphics[scale = 0.15]{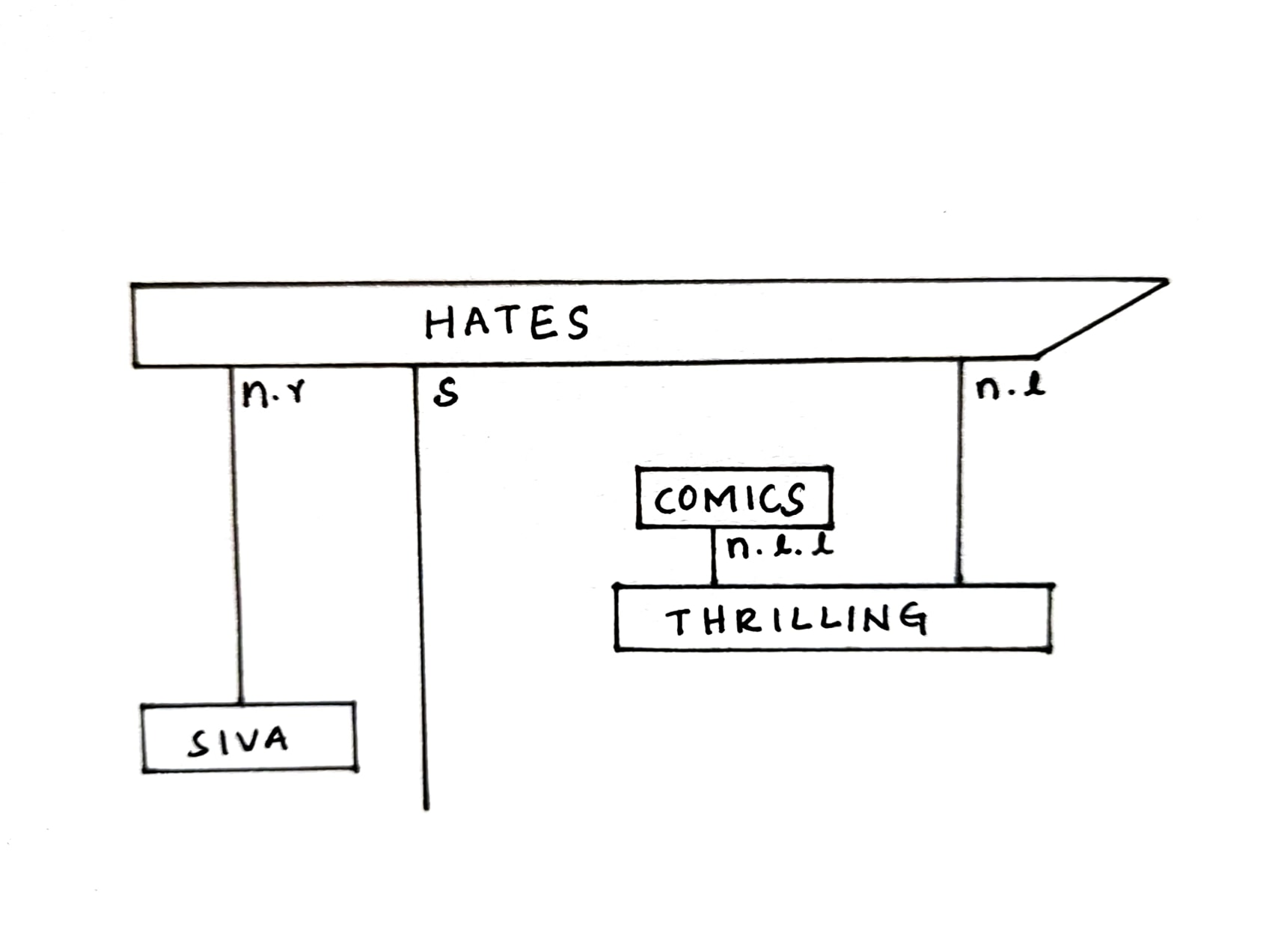}}
\caption{Rewritten string diagram of "siva hates thrilling comics".}
\label{fig2}
\end{figure}

Since cups consume twice the number of qubits assigned to either 'n' or 's', therefore, removing the cups from the original string diagrams results in a diagram which actually consumes less number of resources and is better adapted to today's NISQ hardware. From Fig.~\ref{fig2} it can be seen that the diagram has reduced in size by removing the cups. 

\subsection{Language into Quantum Circuits}\label{languageintocircuit}

After we have designed the string diagrams for the language, we have to transform them into quantum circuits in order to run them on simulators and quantum hardware. The compositional model of meaning described before follows a bottom-up approach i.e. composing words to form the meaning of sentence. On the contrary, language in the form of quantum circuits follows a top-down approach i.e. inferring the meaning of words using the meaning of sentence. This top-down approach is valid because for training quantum circuits a dataset of sentences with labels is provided and from that the meaning of words is inferred. 

\begin{figure}[htbp]
\centerline{\includegraphics[scale = 0.2]{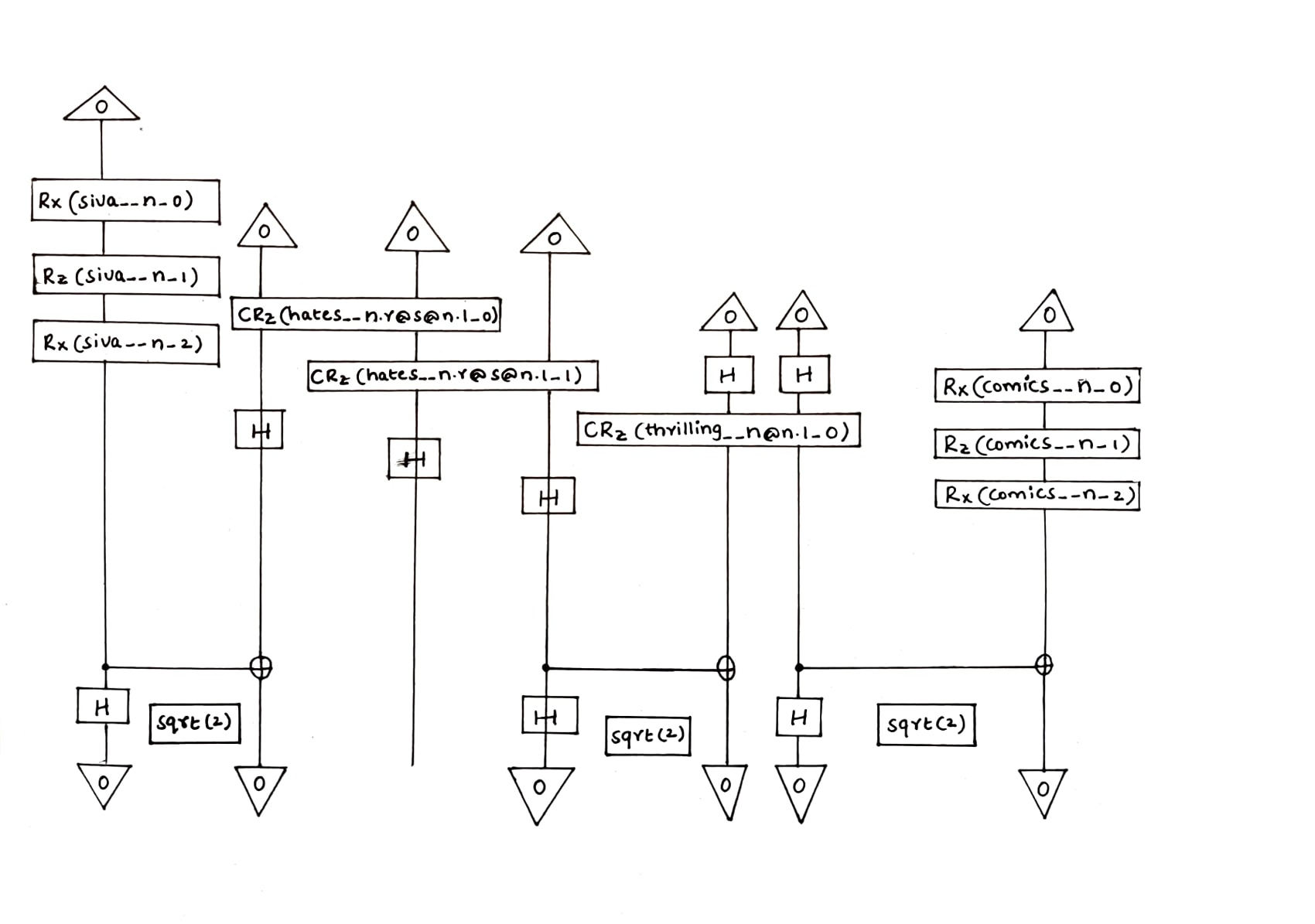}}
\caption{Circuit diagram of "siva hates thrilling comics".}
\label{fig3}
\end{figure}

Fig.~\ref{fig3} refers to the quantum circuit of the string diagram shown in Fig.~\ref{fig1}. The upper triangles (faced upwards) are called "states" or $|states>$ and the lower trianlges (faced downwards) are knows as "effects" or $<effects|$. It can be seen that there are 7 states i.e. 7 qubits and 6 effects i.e. 6 qubits for post-selected measurement. This means that 6 out of 7 qubits need to be used for measurement. 

\begin{figure}[htbp]
\centerline{\includegraphics[scale = 0.16]{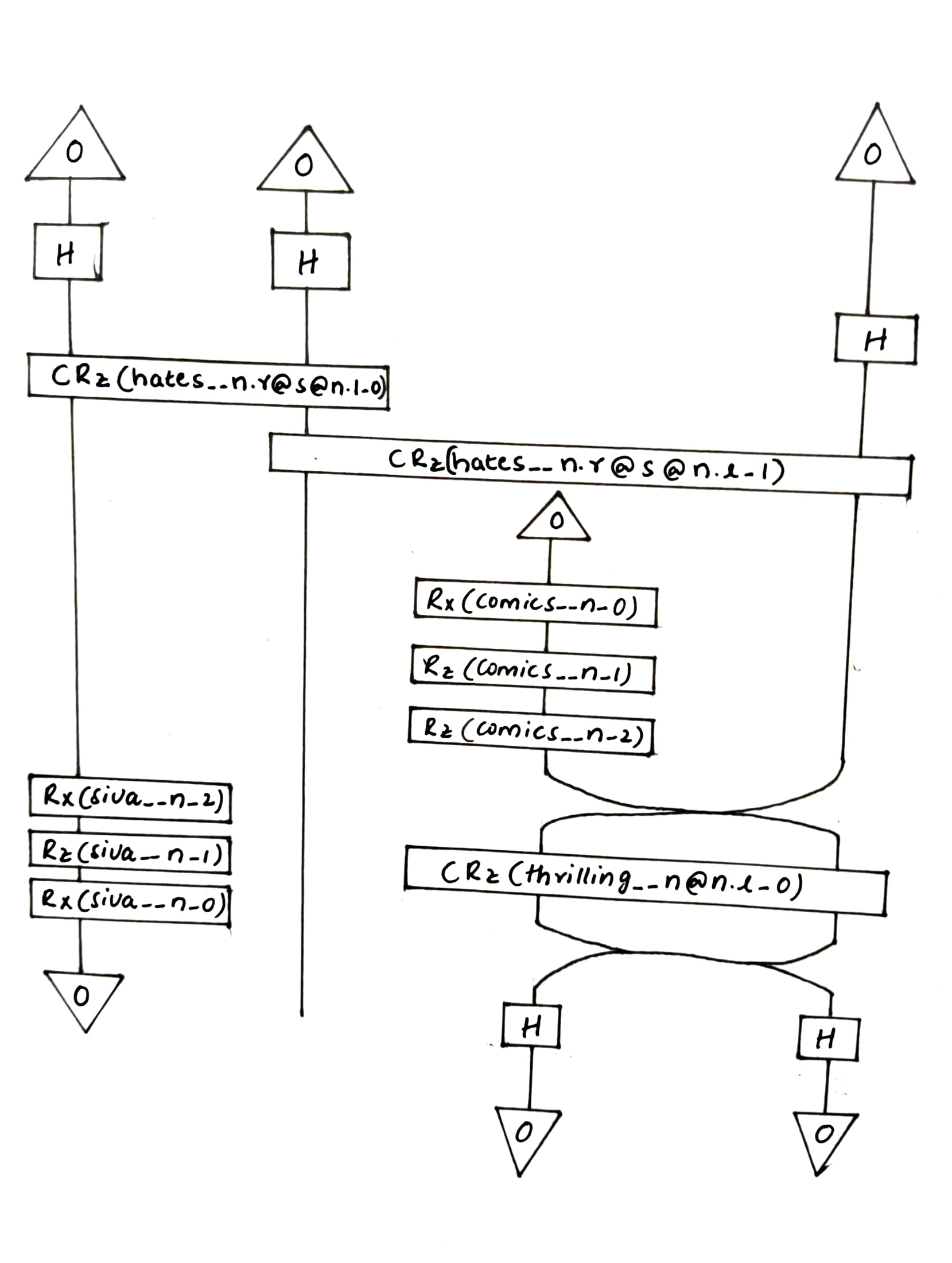}}
\caption{Rewritten circuit diagram of "siva hates thrilling comics".}
\label{fig4}
\end{figure}

Fig.~\ref{fig3} shows that we have Hadamard gates and CNOT gates used to create the entangling effects or cups present in the string diagram. The nouns such as "siva" and "comics" have been converted into circuit form using parametrized quantum gates - $Rz(\alpha)$ and $Rx(\alpha)$. The verbs such as "hates" and "thrilling" have been denoted by parameterized controlled Rz gates. This quantum circuit is denoted as Instantaneous Quantum Polynomial (IQP) \cite{IQPAnsatz} which consists of fixed Hadamard gates, parametrized single qubit gates and controlled two qubit quantum gates. Since the IQP consists of parametrized gates which can vary or modify their output based on input parameters, so it is an example of a variational quantum circuit. 

According to the original string diagram quantum circuit, we require 7 qubit quantum hardware to run the sentence on a quantum computer. Fig.~\ref{fig4} displays the quantum circuit of the diagram in Fig.~\ref{fig2} and it can be deciphered that there are 4 qubits (states) and 3 qubits for post-selected measurement. This is a great reduction of qubits instead of having 6 of 7, we get 3 of 4 after removing the cups by using rewriting technique. Therefore the rewritten circuit can be utilized for NISQ devices.

\subsection{Experimental Details}\label{experimentaldetails}

For conducting sentiment analysis on a quantum computer we have used a binary sentiment classification dataset which contains positive and negative sentiments of candidates on reading various book generes such as fiction, nonfiction, comics and classics. A label of 0 is assigned to positive sentiments and a label of 1 is assigned to negative sentiments. The dataset consists of 130 sentences, out of which 70 are in the training set, 30 in development set and 30 in test set. There are 7 nouns, 3 adjectives and 5 verbs in total for the sentences present in the dataset. 

We have employed lambeq \cite{kartsaklis2021lambeq}, world's first QNLP toolkit for carrying out our experiments. This toolkit provides a convenient way of converting string diagrams into quantum circuits and then using those circuits for each of the sentences for training purpose on a quantum computer. The toolkit itself is based on Python programming language and entails unique features - high level, open source - code available on GitHub, modular - gives independent modules for greater flexibility, extensive - object-oriented design and interoperability - simple communication with other packages.

\begin{figure}[htbp]
\centerline{\includegraphics[scale = 0.4]{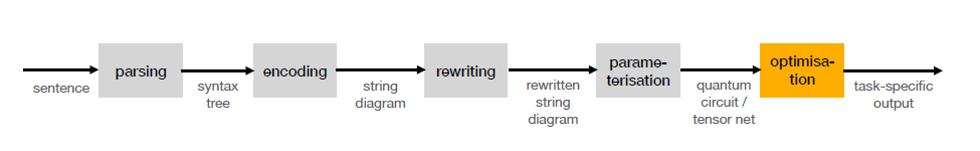}}
\caption{lambeq QNLP Toolkit Pipeline. Fig from \cite{kartsaklis2021lambeq}.}
\label{figlambeq}
\end{figure}

The lambeq pipeline shown in Fig.~\ref{figlambeq} is the general process for QNLP training. A sentence is first parsed by a parser and then converted into a string diagram. Here lambeq uses the state-of-the-art DepCCGParser given in \cite{yoshikawa-etal-2017-astarccg} to parse the sentences in a CCG format and then converts them to string diagrams. The process of converting CCG to string diagram and vice-versa has been explained in \cite{yeung2021ccgbased} by considering CCG as a biclosed category. 

After the sentence is converted into a string diagram, it is converted into a quantum circuit based on the ansatz present in lambeq. There are several ansatz which lambeq provides such as SpiderAnsatz, TensorAnsatz, IQPAnsatz, etc. For each sentence present in the dataset, a circuit is formed and for all the sentences in the dataset those circuits are stored in a list. Based on the optimization scheme chosen, these circuits are sent to the simulator or quantum hardware for training. 

The training process in QNLP is very similar to that of a classical machine learning method. The circuits are ran one by one, measurements are collected from each of the circuits into prediction labels using classical post-processing. These prediction labels are compared with the true labels using a suitable cost function and the output of the cost function is fed into a classical optimizer which calculates the new parameters of the quantum gates. These modified parameters are fed back into the variational circuit again and the process repeats until convergence.


\section{Results and Discussions}

We apply QNLP using lambeq toolkit to our sentiment analysis dataset, and there are four simulation types which we cover:  In classical pipeline the sentences in our dataset are modeled as tensor networks; Quantum pipeline simulation without noise;  Quantum pipeline simulation using JAX,  JAX is a powerful scientific computing library used for automatic differentiation; Quantum pipeline simulation with noise, by using IBM Qiskit’s fake hardware simulator.  We have used FakeVigo as the fake hardware backend simulator for this experiment; it can be changed according to one’s requirement. 

\subsection{Classical Simulation}\label{classicalpipeline}

In all the four experiments, first we convert each sentence present in our data set into string diagrams using the DepCCGParser. Once the sentences are converted into string diagrams, we apply Spider anstaz by which the noun and sentence spaces each receive a dimension of 2. The PyTorch backend is used for the training with Adam optimizer.

\begin{figure}[htbp]
\centerline{\includegraphics[scale = 0.43]{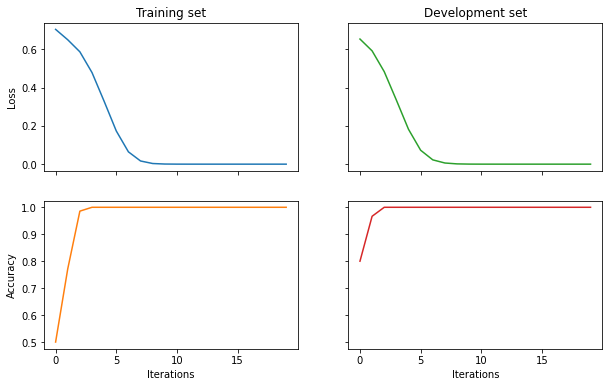}}
\caption{Results of classical pipeline simulation.}
\label{fig5}
\end{figure}

The plots for accuracy obtained on training and development sets are shown in Fig.~\ref{fig5}. We have obtained perfect accuracy on the test set for this case.

\subsection{Noiseless Quantum Simulation}\label{noiselessquantum}

This experiment is not very different from the classical one. We use the same configuration as we used in classical pipeline, however, we change the backend from PyTorch to IBM Qiskit’s Aer simulator that is accessible through pytket, which is $t|ket>$'s python interface. We use a gradient-approximation technique called Simultaneous Perturbation Stochastic Approximation (SPSA) \cite{SPSA}. The reason to choose this optimizer is because SPSA does not calculates the gradient of a quantum circuit but rather approximates it. Evaluating gradients on a quantum hardware by differentiating quantum circuits is very resource intensive and this is where SPSA comes to the rescue.

\begin{figure}[htbp]
\centerline{\includegraphics[scale = 0.43]{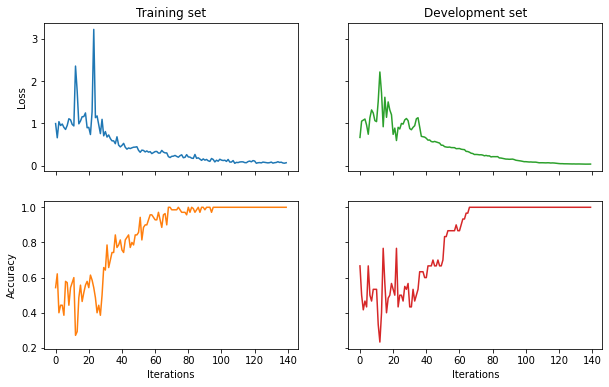}}
\caption{Results of quantum noiseless pipeline simulation.}
\label{fig6}
\end{figure}

Even though there is a lot of instability during the early stages of training as shown in Fig.~\ref{fig6}, but eventually the model converges to good accuracy. The test set accuracy using quantum pipeline is also perfect but the performance varies depending upon on the number of iterations. 

\subsection{Quantum Simulation with JAX}\label{quantumjax}

The string diagrams are changed into variational quantum circuits using the IQP anasatz.  The noun and sentence types get a single qubit each and the layers of IQP are set to 1. We use JAX because the prediction functions are compiled with great speed and JAX takes a very short time to run and execute the results.

\begin{figure}[htbp]
\centerline{\includegraphics[scale = 0.43]{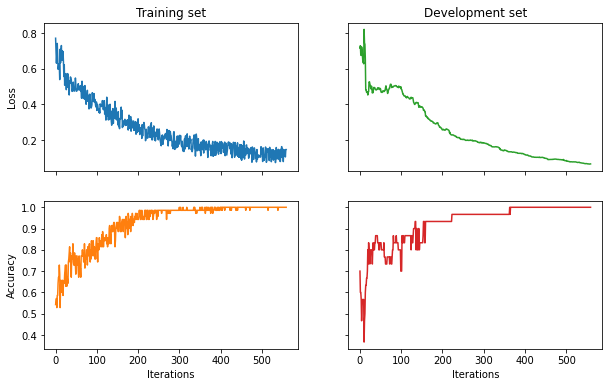}}
\caption{Results of quantum pipeline with JAX simulation.}
\label{fig7}
\end{figure}

Results can be seen in Fig.~\ref{fig7}. Even though we obtain perfect accuracy using JAX, this configuration needs 4 times the number of iterations of noiseless quantum pipeline to gain this feat. 

\subsection{Noisy Quantum Simulation}\label{noisyquantum}

Running circuits on real hardware, taking into account the 130 circuits needed, will be quite difficult and time consuming. So to make this process a bit simpler, we have used Qiskit’s fake quantum hardware backend called FakeVigo with the help of $t|ket>$. The FakeVigo hardware has 5 qubits and is easily able to run our circuits because of the diagram rewriting we have employed. If we exclude the noise, everything else is same in noisy pipeline. 

\begin{figure}[htbp]
\centerline{\includegraphics[scale = 0.43]{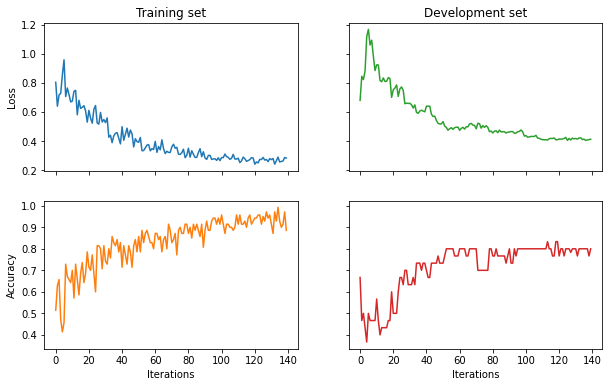}}
\caption{Results of quantum noisy pipeline simulation.}
\label{fig8}
\end{figure}

The test accuracy is not perfect for noisy quantum simulation which can be seen in Fig.~\ref{fig8}. We achieved 83.33\% accuracy on the test set. To attain perfect accuracy it requires even more iterations. To compare and show the difference of how noisy quantum simulation differs from noiseless quantum simulation we haven’t ran the experiment with increased number of iterations. 

\section{Conclusions and Future Work}

In this paper, we have showed the first application of QNLP - binary sentiment classification using the lambeq QNLP toolkit on an intermediate dataset consisting of book genre sentiments. We were able to achieve successful convergence for all the simulations carried out using classical and quantum pipelines. Perfect accuracy on the test set was achieved for three simulations and a decent accuracy was obtained for the noisy quantum pipeline case. 

QNLP is a new field and much work needs to be done in this field in order to achieve quantum advantage. The current work can be extended by including more number of nouns, adjectives and verbs for each of the sentiments in the dataset. This will increase the parameter space. We have performed binary sentiment classification, therefore another direction would be include multi class sentiment classification by including neutral sentiments as well. It would be a great direction of research if random sentences (without following a particular pattern) are also being utilized in the dataset which is of interest to us as that will provide intuition about the scalability aspects of QNLP. 

\section*{Acknowledgment}

S. G. is very grateful to L. M. P. C. for his guidance and suggestions for improvement of the experiments and to Universidad Politécnica de Madrid for supporting this research. S. N. M. thanks Karunya Institute of Technology and Sciences for letting him explore research directions in the field of quantum technology. The authors acknowledge the use of Google Colab Pro for carrying out the experiments and the libraries lambeq \& $t|ket>$ from Cambridge Quantum (Quantinuum). 

\bibliographystyle{./IEEEtran}
\bibliography{./IEEEabrv,./IEEEexample}


\end{document}